\documentclass{pasj00}
\draft

\begin{document}
\SetRunningHead{K. Asano and F. Takahara}
{Cascade Process as a Mechanism of Photon Emission in Gamma-Ray Bursts}
\Received{2002/August/5}%{2002/August/5}
\Accepted{2003/February/20}%{yyyy/mm/dd}

\title{Photon Emission in a Cascade from Relativistic Protons 
Initiated by Residual Thermal Photons in Gamma-Ray Bursts}
\author{Katsuaki \textsc{Asano} and Fumio \textsc{Takahara}}
\affil{Department of Earth and Space Science,
Osaka University, Toyonaka 560-0043}
\email{asano@vega.ess.sci.osaka-u.ac.jp, 
       takahara@vega.ess.sci.osaka-u.ac.jp}


\maketitle

\begin{abstract}
Gamma-ray bursts are generally considered to be the result of 
internal shocks generated in an inhomogeneous relativistic 
outflow that arises from a fireball.
In such shocks, the Fermi acceleration of protons is naturally 
expected to be at least as efficient as that of electrons.
We investigate the consequences of proton acceleration 
in the residual thermal photon field of a fireball,
especially those on the emission spectra of photons.  
In contrast to most other studies, we do not invoke a direct 
electron acceleration in shock waves. 
We show that the residual radiation field of the fireball can 
ignite the photopion production and subsequent cascade, and that 
the photons emitted in this process  
further enhance the photon-initiated cascade.
We find that Fermi accelerated protons with $\gtrsim 10^{13}$ eV 
efficiently bring their energy into pion production and subsequent 
photon emission.
The particles cascading from the pions emit photons over a wide 
energy range. The photons emitted from electron--positron pairs
distribute continuously from the GeV range down to the X-ray range, 
while muons can emit gamma-rays by synchrotron radiation 
with a break at around 1--10 MeV. 
We also discuss several radiation processes which may possibly
produce a break feature in the MeV range, as observed 
in addition to muon synchrotron radiation.
\end{abstract}

%%%%%%%%%%%%%%%%%%%%%%%%%%%%%%%%%%%%%

\section{Introduction}

\indent

A widely accepted scenario for producing gamma-ray bursts (GRBs)
is dissipation of the kinetic energy of a 
relativistic flow by relativistic shocks (see, e.g., a review by 
\cite{pir99}). The rapid time variabilities observed require that 
the GRB itself must arise from internal shocks within the flow, 
while the afterglow is due to external shock produced as the 
flow is decelerated upon collisions with the ambient medium.
Although how the central engine of GRB produces the relativistic flow
remains an open problem, the standard scenario is the high-entropy
fireball scenario \citep{she90,ree92,mes93},
in which a large amount of internal energy
is injected into a compact region, such that the internal
energy per one proton far exceeds the rest mass of the proton \citep{goo86,pac86}.
The huge internal energy of the fireball will be converted
into the bulk kinetic energy of the relativistic flow,
which we call the fireball wind.
This fireball wind is accelerated to a relativistic
velocity and the internal energy of the wind
is finally dominated by baryonic matter.
It is assumed that the fireball wind is inhomogeneous in order to
produce internal shocks.
The inhomogeneity in the wind
may grow, and it ends up with multiple shells with different bulk 
Lorentz factors.

It is to be noted that protons carry a much larger amount of energy 
than electrons in the flow. 
In the standard GRB model, the observed spectra of GRBs are 
interpreted in terms of the synchrotron radiation 
emitted by relativistic electrons, which requires that a large 
fraction of the kinetic energy carried by protons is efficiently 
converted into that of relativistic electrons in the shocked region. 
However, this premise has not been theoretically proven. 
It is apparent that the Coulomb interaction cannot transport the internal 
energy of heated protons 
into electrons to achieve energy equipartition \citep{tot99}, because
the time scale of the Coulomb interaction is much longer
than the dynamical time scale.
As long as protons and electrons are accelerated independently in the 
shocked region, the total energy of accelerated protons should 
overweigh that of accelerated electrons by an order of their mass 
ratio. Although more efficient electron acceleration mechanisms in 
shocks may exist, at present no convincing one has been
specified.
On the other hand, efficient proton 
acceleration may lead to high-energy photon emission through 
photon-initiated cascades, as investigated in this paper. 
Even if we regard that efficient electron acceleration by shocks is 
quite possible, proton acceleration is naturally expected to be at 
least as efficient, and
most probably more efficient; thus, the consequences of proton 
acceleration on photon spectra should be investigated.   

One of the most important characteristics of GRB spectra 
is the existence of a typical break energy scale. 
The observed spectra of GRBs are approximated by a broken power 
law and the photon number spectra are approximated as 
$\sim \varepsilon_\gamma^{-2}$ above the break energy and 
$\sim \varepsilon_\gamma^{-1}$ below that.
The apparent clustering of the break energy of GRB spectra in 
the 50 keV--1 MeV range 
\citep{pre00} is an important key to understanding the 
physical processes in GRBs.
Within the scheme of the standard model, the clustering of the break 
energy and the paucity of low energy photons below the break 
in GRB spectra (Preece et al. 1998, 2000) remain to be explained. 
The break energy scale has been conventionally adjusted by choosing
the minimum energy of accelerated electrons. It is thus interesting 
to see if proton acceleration can provide other mechanisms to 
reproduce the break energy scale. 

Although direct synchrotron emission from protons
is too weak and too soft to reproduce GRBs for the expected values of the 
magnetic field, the emission from secondary particles is not safely 
neglected, provided that 
the Fermi-accelerated protons account for a large fraction of the 
internal energy and that there exist copious target photons to trigger 
the cascades.
In order to optimistically 
estimate the high-energy neutrinos and extremely high-energy cosmic rays 
from GRBs, the Fermi-accelerated protons have often been assumed to have
a large fraction of the internal energy. 
The physical conditions in GRBs imply that
protons may be Fermi-accelerated to high energies \citep{wax95}.
\citet{wax97} pointed out that
a large fraction of the fireball energy
is expected to be converted into high-energy neutrinos
by photomeson production from the accelerated particles.
In this paper, we explore the 
consequences of proton acceleration on the photon spectra and 
investigate if GRBs themselves 
are produced by proton acceleration. 

Assuming that the accelerated protons account for a large fraction of the 
internal energy, we discuss the emission from particles produced 
in the cascades from high-energy protons.
For simplicity and clarity, we assume that the energy density of 
directly accelerated electrons is very small compared to that of protons,  
so that the emission from this component is negligible in contrast 
to that found in almost all previous studies. 
In section 2, we argue that
the residual radiation field of the fireball
can trigger photopion production of relativistic protons.
In section 3, the emission
from the cascading particles is investigated.
In section 4, we discuss the possibility to explain
the observed spectral property of GRBs.
Our conclusions are summarized in section 5.

\section{Cascade Initiated by a Residual Radiation Field of the Fireball}

\indent

In this section we discuss photopion production in the internal shock
against residual thermal photons of the fireball.
While most of previous studies have ignored the effects of the residual photons, 
we argue that this is not legitimate, and that these photons play an 
important role to trigger photopion production.
The cascade triggered by the collisions of shock-accelerated protons 
with residual photons can efficiently transport the dissipated energy 
into electron--positron pairs.
A schematic diagram of the energy transfer is shown in figure 1.
In the following, we discuss the ignition of GRBs 
according to this diagram.

\begin{figure}
\begin{center}
\FigureFile(100mm,170mm){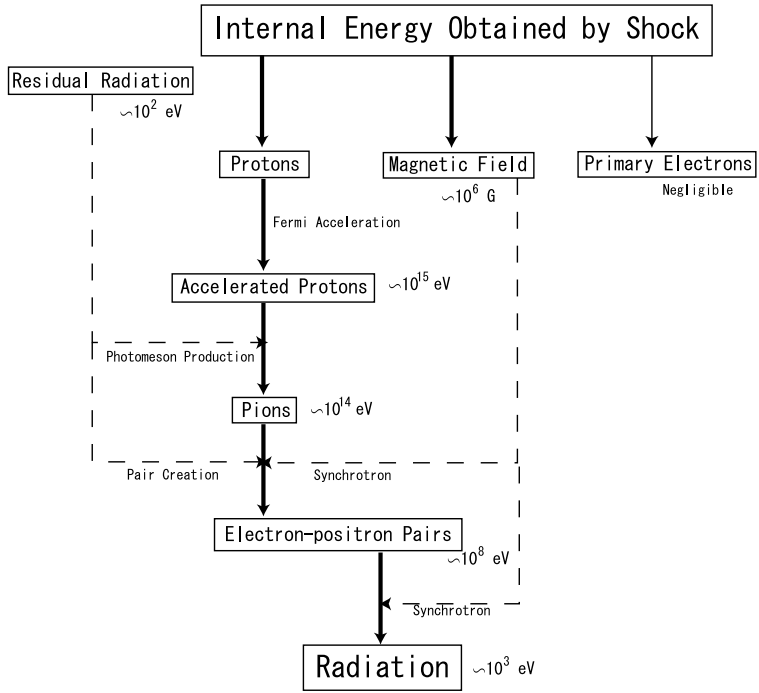}
\end{center}
\caption{Process of the cascade ignited by
the residual radiation in the fireball (see section 2).
We neglect the effects of direct electron acceleration. 
The solid arrows show the energy flow.
The dashed arrows indicate the interactions that induce energy transfer
to lower energy particles.
The typical energies of each particle measured in the comoving frame
are noted.
}
\end{figure}

In order to discuss this issue, 
we need to specify the distribution of the residual radiation 
field. We regard that the central engine produces an inhomogeneous 
fireball wind for a long duration compared to the dynamical time 
scale of the central engine. The inner region of the fireball wind 
is optically thick and the radiation and matter in the fireball 
wind behaves like a single fluid moving with the same velocity.
In the radiation-dominated stages, as the wind expands, the Lorentz 
factor of the wind increases in proportion to the radius \citep{pir93}.
After the fireball wind becomes matter-dominated,
the wind plasma coasts with a constant radial speed.
The fireball wind becomes optically thin at the photospheric radius
where matter and radiation decouple. 
We treat this decoupled radiation field as the residual radiation field.

In the internal shock scenario, there exist inhomogeneities of the matter 
in the wind, and they end up with multiple shells with different bulk 
Lorentz factors. Internal shocks arise from collisions of the shells.
As for the Lorentz factor of the shells, we follow the general 
constraints obtained in the conventional internal-shock scenario. 
The bulk Lorentz factor of the shells lies between one hundred and 
a few thousand \citep{mes00}. Moreover, the energy-efficiency 
argument concerning the internal shock model requires a large dispersion 
of the Lorentz factor of individual shells \citep{kob01,gue01}.
Efficient dissipation of the kinetic energy is possible only when 
the difference in the Lorentz factors of the two colliding shells 
is sufficiently large. Thus, we regard that independent multiple 
shells are arising intermittently from the central engine
to satisfy the required large dispersion of their Lorentz factors.
 
On the other hand, the inhomogeneity of the residual radiation field 
is less clear and the degree of inhomogeneities largely depends on 
the character of the central engine, about which we have little information.
The residual photon field in the wind may not necessarily distribute
discontinuously as do the matter shells.
Therefore, we consider two extreme cases for the residual photon field
(see figure 2).
In one case, we assume that radiation is strongly coupled with matter 
and is contained in the shell. In this case, each shell is regarded to 
represent an independent fireball.
We assume that there is no matter and radiation between the shells before
the radiation field in a shell decouples with matter.
In this case, we completely neglect the radiation field emitted from
other shells.
Hereafter, we call this case the ``Discrete Model''.

In the other extreme case, we assume that the photon field is given by
a quasi-steady continuous wind, for which the effects of
the inhomogeneities are neglected.
Inside the photosphere the radiation and matter are considered
to be continuously distributed and coupled with each other.
Outside of the photosphere photons escape freely to infinity.
In this case the radiation field behaves smoothly,
as determined from the wind parameters. 
Even though this treatment formally neglects the inhomogeneities in the wind,
in reality inhomogeneities should exist, and
we assume that the internal shocks occur and high-energy protons
are injected at some places in the radiation field.
We do not discuss the detail of the internal shocks in this model.
The latter case is called  the ``Continuous Model'' hereafter.

For simplicity, we assume spherical symmetry for both models. 

\begin{figure}
\begin{center}
\FigureFile(80mm,80mm){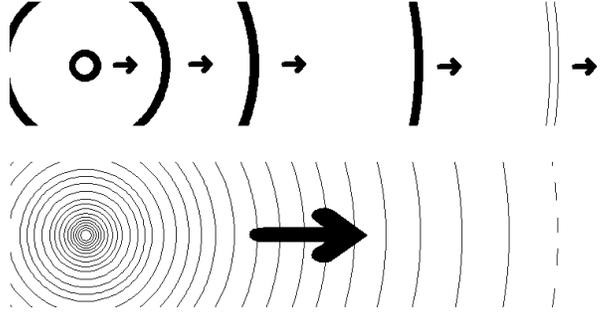}
\end{center}
\caption{Schematic representation
for the Discrete Model (above) and the Continuous Model (below).
The white shell (above) and the dashed lines (below)
represent fluid outside of $R_{\rm ph}$.}
\end{figure}

\subsection{Residual Photon Field}

\subsubsection{Discrete Model}

We first discuss the Discrete Model. 
We consider the situation where the inner faster shell overtakes 
the outer slower one and generates 
forward and reverse shocks in the respective shells.
We suppose that the typical Lorentz factors of the faster and slower 
shells are $\Gamma_{\rm r}\sim $ 1000 and $\Gamma_{\rm s}\sim $ 100, 
respectively.
Each shell is emitted from the central engine as a fireball
with some initial radius $\Delta$.
Each fireball ends up as a coasting matter shell, retaining
its width ($\Delta$) in the observer frame.
The radiation field in each shell evolves independently of other 
shells. In the initial radiation-dominated phase,
the radiation energy density in the fireball behaves as 
$e \propto r^{-4}$ ($r$ is the location of the shell),
and the bulk Lorentz factor of the shell
increases in proportion to $r$ \citep{pir93}.
After the shell enters the matter-dominated regime, 
as long as the shell is optically thick to scattering, the
radiation energy density declines as $e \propto r^{-8/3}$ and 
the shell eventually coasts with $\Gamma = E/Mc^2$,
where $M$ and $E$ are the total baryon rest mass and the total energy 
of the shell, respectively.
Since there is no matter between shells in the Discrete Model,
once a photon escapes from the shell, it moves freely.

In this case, the shell becomes optically thin at the radius
%%%%%%%%%%%%%%%%%%%%%%%
\begin{eqnarray}
R_{\rm ph}=\sqrt{\frac{\sigma_{\rm T} M}{4 \pi m_{\rm p}}},
\label{photo}
\end{eqnarray}
%%%%%%%%%%%%%%%%%%%%%%%
where $\sigma_{\rm T}$ is the Thomson cross section.
The radius $R_{\rm ph}$ becomes
$6 \times 10^{12}$--$6 \times 10^{13}$ cm
for a range $Mc^2=10^{48}$--$10^{50}$ erg.
The kinetic energy of one shell which we consider corresponds to
$10^{50}$--$10^{53}$ erg for $\Gamma =100$--$1000$.
Then, we obtain the temperature of the residual radiation field
at $r=R_{\rm ph}$ in the shell rest frame as
%%%%%%%%%%%%%%%%%%%%%%%
\begin{eqnarray}
T=130 (\Gamma_3 \Delta_7 M_{48})^{-1/12} \quad \mbox{eV},
\label{temp}
\end{eqnarray}
%%%%%%%%%%%%%%%%%%%%%%%
where $\Gamma_3=\Gamma/10^3$, $\Delta_7=\Delta/10^7$ cm,
and $M_{48}=M c^2/ 10^{48}$ erg.
The parameter dependence of the temperature is very weak.
The shell coasts $R_{\rm esc} \equiv 2 \Gamma^2 \Delta
\simeq 2 \times 10^{13} \Delta_7 \Gamma_3^2$ cm from the radius 
$r=R_{\rm ph}$ before the residual photons escape from the shell. 
Thus, the energy density of residual photons in the comoving frame 
is as large as $10^{10} \mbox{ erg ${\rm cm^{-3}}$}$ when the shell 
reaches a few times $10^{13}{\rm cm}$ for $\Gamma=10^3$.  

The inner faster shell overtakes the outer slower one at
$r \simeq \Gamma_{\rm s}^2 \delta$, where $\delta$ is the initial 
separation between the two shells in the observer rest frame. 
When $\delta$ is small enough, the collisions occur well inside of
$R_{\rm ph}$ and no major observational feature is expected. 
For $\delta \simeq 10^9$ cm (which is consistent with the observed time variability
of GRBs) and $\Gamma_{\rm s} \gtrsim 100$, 
the two shells collide outside of $R_{\rm ph} \simeq 10^{13}$ cm.
The typical radius where the internal shocks occur can be comparable 
to $R_{\rm ph}$,
unless the initial separation of two shells is limited to larger 
than $10^{10}$ cm.
Since the collision radius is comparable to $R_{\rm esc}$
for the faster shell with $\Gamma_{\rm r} \simeq 10^3$,
internal shocks can occur in the thermal radiation
field in the faster shell with temperature $T \simeq 100$ eV.
On the other hand, when $\delta$ is much larger than $10^{10}$cm,
internal shocks occur far outside of $R_{\rm ph}$ and the residual
photons have already escaped from the shell.
In this case there is no photon field in the shell when the internal 
shocks occur, which is not interesting for the present study.
In the Discrete Model, therefore, we consider only the internal 
shocks that occur around the radius $r=R_{\rm ph}$.
In the comoving frame the width of the shocked region in 
the outer shell is thinner than that of the inner shell when 
$\Delta$ and $M$ of the two shells are similar to each other. 
Since the photons in the shocked region in the outer shell 
escape more promptly,
we discuss only the reverse shocked region of the inner shell.
Hereafter, we denote $\Gamma_3$ as the normalized Lorentz factor 
of the inner shell ($\Gamma_{\rm r}=10^3 \Gamma_3$).

The internal shock is mildly relativistic, i.e.,  
the Lorentz factor $\Gamma_{\rm R}$ of the reverse shock
observed from the rest frame of the inner shell is a few.
For example, when two shells with the same $M$ and $\Delta$ collide
with $\Gamma_{\rm s}=100$ and $\Gamma_{\rm r}=1000$,
we have $\Gamma_{\rm R}=2.57$.
From the jump condition \citep{sar95},
the shell width in the comoving frame of the shocked matter is
$l=\Delta \Gamma_{\rm r}/(4 \Gamma_{\rm R}+3) \simeq 
10^9 \Delta_7 \Gamma_3$ cm.
Then, the internal energy density in the shocked shell at 
$r=R_{\rm ph}$ is obtained as
%%%%%%%%%%%%%%%%%%%%%%%
\begin{equation}
U_{\rm in}(R_{\rm ph}) = \frac{(\Gamma_{\rm R}-1) M c^2}
{4 \pi R_{\rm ph}^2 l}
=\frac{(\Gamma_{\rm R}-1) (4 \Gamma_{\rm R}+3) m_{\rm p} c^2}
{\sigma_{\rm T} \Gamma_{\rm r} \Delta}
\simeq 10^{12} \Delta_7^{-1} \Gamma_3^{-1} \quad 
\mbox{ erg ${\rm cm^{-3}}$},
\end{equation}
%%%%%%%%%%%%%%%%%%%%%%%
where we have assumed 
$(\Gamma_{\rm R}-1) (4 \Gamma_{\rm R}+3) \simeq 10$.
Thus, the energy density of residual photons is about 
1\% of the internal energy of baryons. 

In the internal or external shocks of GRBs
strong magnetic fields should be produced,
since the expected magnetic field frozen from the central engine
is several orders of magnitude smaller than the value required
to produce gamma-rays because of the adiabatic expansion.
Although no specific mechanism is widely accepted,
the relativistic two-stream instability \citep{med99}
is one of the candidates to generate such a high magnetic field.
In this paper, we assume that the constant ratio ($f_{\rm B}$) of
the internal energy ($U_{\rm in}$) goes into the magnetic field;
then, the magnetic field at radius $R_{\rm ph}$ becomes
%%%%%%%%%%%%%%%%%%%%%%%
\begin{equation}
B(R_{\rm ph}) =\sqrt{8 \pi f_{\rm B} U_{\rm in}}
\simeq 2 \times 10^6 \left( \frac{f_{\rm B}}{0.1} \right)^{1/2}
\Delta_7^{-1/2} \Gamma_3^{-1/2} \quad {\rm G}.
\end{equation}
%%%%%%%%%%%%%%%%%%%%%%%

\subsubsection{Continuous Model}

Differently from the Discrete Model, for the Continuous Model 
we consider a quasi-steady fireball wind
of total luminosity $L=10^{52} L_{52}$ erg ${\rm s^{-1}}$
expanding from the initial radius $\Delta$.
Although the radiation and matter are assumed to be
continuously distributed in the estimation of the distribution
of the physical quantities,
in reality build-in inhomogeneities can make
the internal shocks occur somewhere and high-energy protons
are injected.
We do not deal with the details of the internal shocks in this model.
The injected high-energy protons will interact with
the continuous radiation field.
The wind is characterized by the average baryon load parameter, 
$\eta \equiv L/\dot{M} c^2=100 \eta_2$.
The photosphere and photospheric temperatures in this case
were well studied by \citet{mes00}.
In the radiation-dominated region
the bulk Lorentz factor of the wind increases as $\Gamma \propto r$
($r$ is the radial coordinate in this model).
For relatively low values of 
$\eta < \eta_* \equiv ( L \sigma_{\rm T}/4 \pi m_{\rm p}
c^3 \Delta)^{1/4}=10^3 L_{52}^{1/4} \Delta_7^{-1/4}$, 
the increase of $\Gamma$ stops at the saturation
radius, $r_{\rm s} \equiv \eta \Delta$, and the flow continues to coast with
$\Gamma=\eta$ above $r_{\rm s}$. 
In this paper we consider only this range of $\eta$. 
The photosphere is located at a radius
%%%%%%%%%%%%%%%%%%%%%%%
\begin{eqnarray}
R_{\rm ph}=\frac{L \sigma_{\rm T}}{4 \pi m_{\rm p} c^3 \Gamma^3}
\simeq 10^{13} L_{52} \eta_2^{-3} \quad \mbox{cm}.
\label{photo2}
\end{eqnarray}
%%%%%%%%%%%%%%%%%%%%%%%
We obtain the radiation temperature at $r=R_{\rm ph}$ in the wind rest frame as
%%%%%%%%%%%%%%%%%%%%%%%
\begin{eqnarray}
T=40 \eta_2^{5/3} L_{52}^{-5/12} \Delta_7^{1/6} \quad \mbox{eV}.
\label{temp2}
\end{eqnarray}
%%%%%%%%%%%%%%%%%%%%%%%
Although the representative numerical values of $R_{\rm ph}$ and $T$ 
are similar to those of the Discrete Model, their parameter dependence 
is rather strong compared with the dependence in the Discrete Model.
Outside of $R_{\rm ph}$, the radiation field is approximated by a diluted 
black body with temperature $T$.
In the wind rest frame, the number density of the residual photons 
declines as $r^{-2}$. Although the energy density of residual photons is 
an order of magnitude smaller than that in the Discrete Model, the photon
distribution is not limited in narrow shells, and they are available 
for a longer time scale. 

The average total energy density in the wind rest frame,
$U = L/4 \pi r^2 \eta^2 c$, is given by
%%%%%%%%%%%%%%%%%%%%%%%
\begin{eqnarray}
U&=&\frac{4 \pi m_{\rm p}^2 c^5 \eta^4}{L \sigma_{\rm T}^2} 
\left( \frac{r}{R_{\rm ph}}
\right)^{-2} \\
&\simeq& 2 \times 10^{10} \eta_2^4 L_{52}^{-1} \mathcal{R}^{-2} 
\mbox{erg ${\rm cm^{-3}}$},
\label{U}
\end{eqnarray}
%%%%%%%%%%%%%%%%%%%%%%%
where $\mathcal{R} \equiv r/R_{\rm ph}$.
If the fraction $f_{\rm B}$ of
$U$ goes into the magnetic field,
it becomes
%%%%%%%%%%%%%%%%%%%%%%%
\begin{equation}
B=2 \times 10^5 \left( \frac{f_{\rm B}}{0.1} \right)^{1/2} \eta_2^2
L_{52}^{-1/2} \mathcal{R}^{-1} \quad \mbox{G}.
\label{B}
\end{equation}
%%%%%%%%%%%%%%%%%%%%%%%
These values are for a continuous wind, and the effects of inhomogeneities are 
not taken into account. When an inhomogeneity causes an internal shock 
which propagates in the background wind flow, 
the internal energy and magnetic field can be larger than the values 
given in equations (\ref{U}) and (\ref{B}) locally.
However, for simplicity, in the following numerical estimation 
we do not consider such factors in the Continuous Model.
The dynamical time scale in the Continuous Model is taken to be 
$r/c\eta$, which is typically two orders of magnitude longer than 
that in the Discrete Model $l/c$. 

\subsection{Photopion Production by Accelerated Protons}

The maximum energy of accelerated protons is estimated 
from the conditions that the cooling time of protons should be 
longer than the Fermi acceleration time scale,
and that the Larmor radius should be smaller than the shell width.
For the Discrete Model, the maximum energy of protons at 
$r=R_{\rm ph}$ is most likely determined by
the latter condition when $f_{\rm B} \ll 1$.
For $f_{\rm B}=0.1$ and $l=10^9$ cm, however, both of the conditions
give comparable values of the maximum energy, which turn out to be 
$\sim 10^{17}$ eV. For the Continuous Model, a similar value of the 
maximum energy is obtained. 
The accelerated protons would obey a power-law distribution and 
may account for a large fraction of the internal energy.
The thermal photons interact with these protons to produce photopions.

Although in the shocked fluid rest frame the residual photons 
undergo a red-/blue-shift and their distribution becomes anisotropic, 
an isotropic distribution is a quite good approximation 
in a relativistic flow.
Thus, we neglect the anisotropy and suppose that
the accelerated protons interact with the isotropic
thermal radiation field with a given temperature.
The typical Lorentz factor of accelerated protons
that interact resonantly with the thermal photons is
$\gamma_{\rm p,typ}=\varepsilon_{\rm th}/T$, where
$\varepsilon_{\rm th} \simeq 145$ MeV is the threshold energy of 
a photon in the proton rest frame.

\subsubsection{Discrete Model}

For the Discrete Model, we consider the internal shocks that
occur around only the radius $r=R_{\rm ph}$.
The typical energy of pion producing protons becomes 
$E_{\rm p,typ}\simeq 10^{15}$eV for $T=100$ eV.
At $r=R_{\rm ph}$ the number density of the photons is written as
%%%%%%%%%%%%%%%%%%%%%%%
\begin{eqnarray}
n(\varepsilon_\gamma)=\frac{8 \pi}{(h c)^3} \frac{\varepsilon_\gamma^2}
{\exp{(\varepsilon_\gamma/T)}-1}.
\end{eqnarray}
%%%%%%%%%%%%%%%%%%%%%%%
The time scale of photopion production, $t_{\pi}$,
is written as
%%%%%%%%%%%%%%%%%%%%%%%
\begin{eqnarray}
t_{\pi}^{-1}(\gamma_{\rm p})=2 \pi c \int_0^\pi d\theta \sin{\theta} 
(1-\cos{\theta})
\int_{\varepsilon'_{\rm th}}^\infty d\varepsilon_\gamma
\frac{n(\varepsilon_\gamma)}{4 \pi}
\sigma_\pi(\chi),
\label{tau}
\end{eqnarray}
%%%%%%%%%%%%%%%%%%%%%%%
where $\theta$ is the incident angle between the two interacting particles.
We approximate the photopion production cross section
(e.g. \cite{ste68}) by a broken power-law profile as
$\sigma_{\pi}(\chi) =5 \times 10^{-28} (\chi/590)^{3.2} {\rm cm^2}$ 
for $290<\chi<590$ and 
$\sigma_{\pi}(\chi) =5 \times 10^{-28} (\chi/590)^{-0.7} {\rm cm^2}$ 
for $590<\chi<9800$, where $\chi m_{\rm e} c^2$
is the photon energy in the proton rest frame.
In the fluid rest frame the energy of a photon is expressed as 
$\varepsilon_\gamma=\chi m_{\rm e} c^2/\gamma_{\rm p} (1-\cos{\theta})$.
Since the residual photons escape in the dynamical time scale $l/c$, 
we regard that efficient pion production occurs if $t_\pi$ is less than $l/c$.
In figure 3, we plot the time scale $t_\pi$ and 
$\tau_{\pi} \equiv t_{\pi}^{-1} l/c$ for the photopion production
in the 100 eV radiation field.
It is seen that for $E_{\rm p}=E_{\rm p,typ}\simeq 10^{15}$ eV
$\tau_{\pi}$ becomes $\sim 8 (l/10^9 \mbox{cm})$, which means that 
efficient pion production occurs. 
The threshold energy of protons for pion production, $E_{\rm p,\tau}$, 
at which $\tau_{\pi}$ becomes one, is $2 \times 10^{14}$ eV for 
$T=100$ eV. Therefore, protons with $10^{14}$--$10^{17}$ eV efficiently 
produce $\pi^0$s and $\pi^\pm$s.
The photopion production would not occur outside of a few times $R_{\rm ph}$
because the photon number density decreases as $\tau_{\pi} \propto r^{-2}$,
even if the residual photon field remains in the shell.
Since the average inelasticity is about 20\%,
the typical energy of produced pions should be $10^{13}$--$10^{16}$ eV.

\begin{figure}
\begin{center}
\FigureFile(100mm,170mm){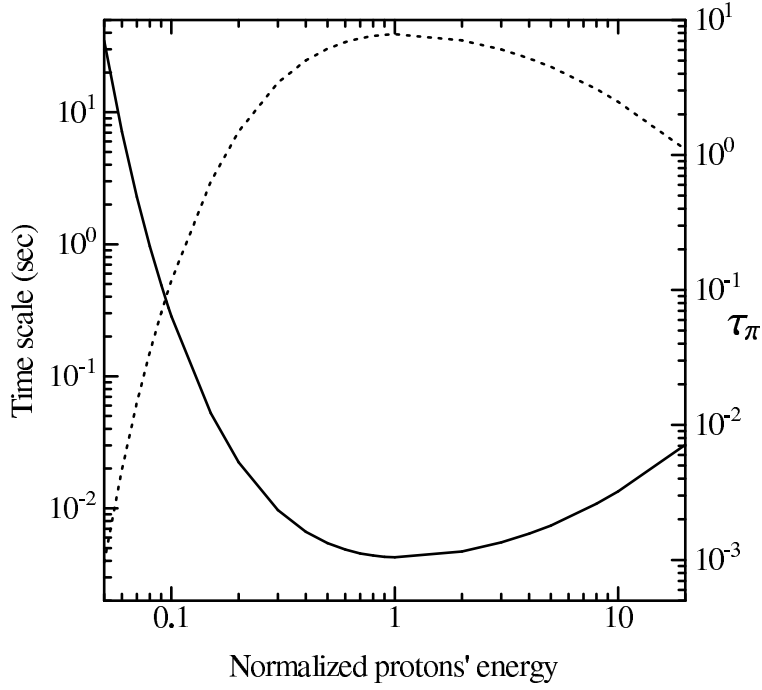}
\end{center}
\caption{Time scale (solid line) and $\tau_{\pi}$ (dotted line)
of protons for photopion production in the thermal photon field.
The time scale $t_{\rm p,\pi} \propto T^{-3}$ and the optical depth
$\tau_{\pi} \propto t_{\rm p,\pi}^{-1} l$.
We assume $T=100$ eV and $l=10^9$ cm in this figure.
The protons' energy is normalized by $E_{\rm p,typ}=m_{\rm p} c^2 \varepsilon_{\rm th}/T$.
}
\end{figure}

\subsubsection{Continuous Model}

When an internal shock occurs in a continuous wind,
the accelerated protons will be injected into the wind.
The accelerated protons can interact with photons during much 
longer period of time ($r/c\eta$) than in the Discrete Model ($l/c$).
In figure 4 we plot the threshold energy, $E_{\rm p,\tau}$, at the photosphere
calculated numerically in a similar way to that in the Discrete Model.
The energy $E_{\rm p,\tau}$ weakly depends on $L_{52}$.
The dotted line in figure 4 corresponds to $\eta=\eta_*$, the maximum 
value allowed for the saturated fireball wind.
For fiducial values of $\eta_2=1$ and $L_{52}=1$, the threshold 
energy becomes $3 \times 10^{14}$ eV, which is almost the 
same as the typical value for the Discrete Model. 
Thus, in the Continuous Model, too, efficient photopion 
production occurs around the photosphere. 

\begin{figure}
\begin{center}
\FigureFile(120mm,80mm){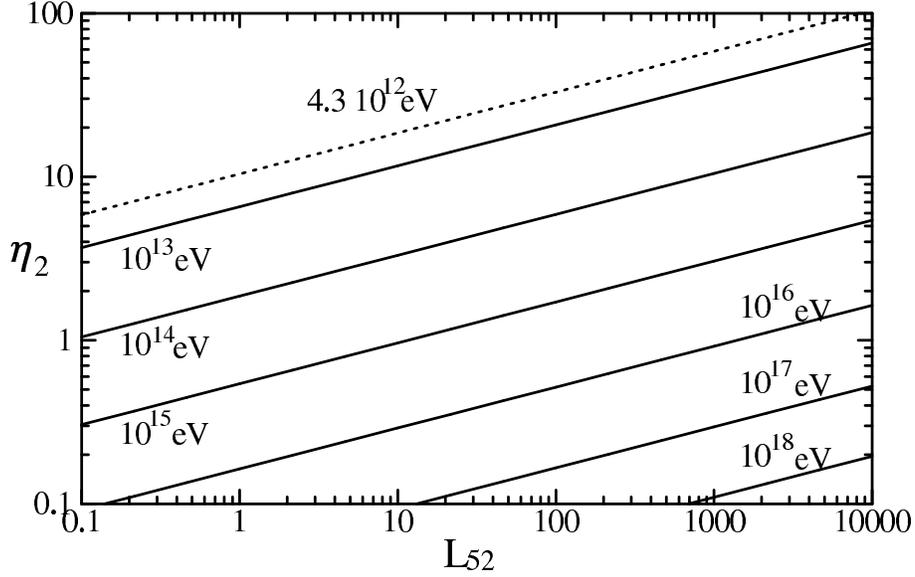}
\end{center}
\caption{Threshold energy of the photopion production,
$E_{\rm p, \tau}$, at the photosphere in the Continuous Case for $\Delta_7=1$.
The dotted line corresponds to $\eta=\eta_*$.
}
\end{figure}

This may be understood qualitatively as follows. Because 
$\tau_{\pi}$ is a steeply increasing function of the energy 
in the low-energy range, as can be seen in figure 3, 
the threshold energy is roughly taken as a fixed fraction of 
the typical energy for photopion production, which is inversely 
proportional to the photon temperature. 
Since $\tau_{\pi}(E_{\rm p,typ}) \propto T^3 \cdot R_{\rm ph}/\eta$, 
we may approximate $E_{\rm p,\tau} \propto E_{\rm p,typ} \propto 
\eta_2^{-5/3} L_{52}^{5/12}$.
It is seen that this simple relation well reproduces the plots in 
figure 4.

Since in the Continuous Model the residual photons distribute 
continuously, the photopion production can occur even outside of 
$R_{\rm ph}$. In figure 5, we show the threshold energy, 
$E_{\rm p,\tau}$, against $r$ for $\eta_2=1$.
Although the value $\tau_{\pi}$ drops as $\propto r^{-1}$,
$E_{\rm p,\tau}$ itself does not increase drastically.
%The photopion production do not occur efficiently outside of
%a radius where $\tau_{\pi}=1$ for protons with $E_{\rm p}=E_{\rm p,typ}$.
The region where efficient photopion production can occur by the thermal 
radiation field is inside of $r \simeq 100 R_{\rm ph}$.
Therefore, we expect a large fraction of the energy of protons 
($\gtrsim 10^{14}$ eV for $L_{52} \simeq 1$) to efficiently go to the energy of 
pions.

\begin{figure}
\begin{center}
\FigureFile(120mm,80mm){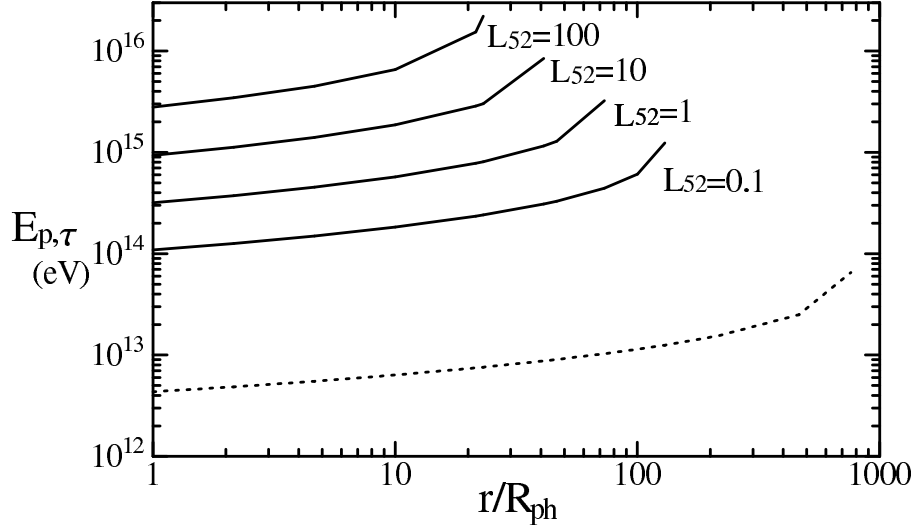}
\end{center}
\caption{Threshold energy, $E_{\rm p, \tau}$, for $\eta_2=1$ and $\Delta_7=1$
against $r$ in the Continuous Case.
The dotted line is the same as in figure 4.
}
\end{figure}

In the Discrete Model we have assumed that the internal shocks occur near 
$r=R_{\rm ph}$. Therefore, the physical conditions in the shocked region 
are somewhat restricted.
On the other hand, in the Continuous Model,
photopion production occurs over a wide range of $r$
and the properties of the photon emission  may depend on
the parameters $\eta$, $L$, and $\mathcal{R}$.
For simplicity, we describe the emission from the cascading particles
mainly along the Discrete Model below, while the Continuous Model is briefly 
mentioned.
As will be shown later, however,
there are no drastic differences in the results between the two Models.

\subsection{Behavior of Pions}

The created $\pi^0$s decay immediately into two gamma-rays, which 
are converted into electron--positron pairs, while the minor portion 
directly escapes from the shocked region. The fate of these 
high-energy photons is discussed in the next subsection. 
On the other hand, the lifetime of most $\pi^\pm$s 
($E_\pi \gtrsim 0.2 E_{\rm p,typ} \simeq 2 \times 10^{14}$ eV)
is longer than the dynamical time 
scale $l/c \simeq 0.03$ s for the Discrete Model 
because of the large Lorentz factor.
The energy density of residual photons is about 
$10^{10} \mbox{ erg ${\rm cm^{-3}}$}$, 
which is smaller than the energy density of the magnetic field 
if $B$ is larger than $5\times 10^5$G. 
Thus, synchrotron loss is the dominant cooling process for 
the charged pions. 

The synchrotron cooling time scale of $\pi^\pm$ is expressed as
%%%%%%%%%%%%%%%%%%%%%%%
\begin{equation}
t_{\rm \pi,cool}=\frac{6 \pi m_{\pi}^3 c}
{\sigma_{\rm T} m_{\rm e}^2 B^2 \gamma_{\pi}}
\simeq 2 \times 10^{-3} \left(
\frac{f_{\rm B}}{0.1} \right)^{-1} \left(
\frac{E_{\pi}}{2 \times 10^{14} \mbox{eV}} \right)^{-1}
\Delta_7 \Gamma_3 \quad \mbox{sec},
\end{equation}
%%%%%%%%%%%%%%%%%%%%%%%
where $\gamma_\pi$ is the Lorentz factor of pions.
The cooling time of high-energy pions 
becomes shorter than their lifetime.
The lifetime of pions becomes comparable to the cooling time
for 
%%%%%%%%%%%%%%%%%%%%%%%
\begin{eqnarray}
\gamma_\pi^2
=\frac{6 \pi m_\pi^3 c}{\sigma_{\rm T} m_{\rm e}^2 B^2 t_{\pi,0}}
\equiv \gamma_{\pi 0}^2,
\end{eqnarray}
%%%%%%%%%%%%%%%%%%%%%%%
where $t_{\pi,0}=2.6 \times 10^{-8}$ s is the lifetime of pions 
in the rest frame.
Thus, charged pions ($E_{\pi} \gtrsim 2 \times 10^{14}$ eV)
cool down to $m_\pi c^2 \gamma_{\pi 0} =
5 \times 10^{13} (f_{\rm B}/0.1)^{-1/2} $eV before they  
decay into muons. Charged pions with lower energies decay into 
charged muons before significant cooling, and most of their 
energy is converted into photons by synchrotron radiation 
because the lifetime of charged muons is much longer than that 
of charged pions. 

The energy of the synchrotron photons
emitted by charged pions is
%%%%%%%%%%%%%%%%%%%%%%%
\begin{equation}
\varepsilon_{\pi}=\frac{\hbar q B}{m_{\pi} c} \gamma_{\pi}^2
\sim 9 \times 10^7 \left( \frac{E_{\pi}}{2 \times 10^{14} \mbox{eV}}
\right)^2 \left( \frac{B}{10^6 \mbox{G}} \right)  \quad \mbox{eV}.
\label{pi-syn}
\end{equation}
%%%%%%%%%%%%%%%%%%%%%%%
Thus, pions of $10^{14}$--$10^{15}$ eV emit
photons of $10^7$--$10^9$ eV. The energy of muon synchrotron radiation 
is distributed mainly in the range of $10^5$--$10^7$eV. 
The conversion efficiency into high-energy photons in the  
cascade process is fairly large, much larger than that into neutrinos. 

In the Continuous Model, since the magnetic field is weaker 
the synchrotron cooling time scale becomes longer, comparable to the 
decay time for charged pions with energy of $10^{15}$eV.
In a similar way to that in the Discrete Model we obtain
$m_\pi c^2 \gamma_{\pi 0} \simeq
5 \times 10^{14} (f_{\rm B}/0.1)^{-1/2} \eta_2^{-2} L_{52}^{-1/2} \mathcal{R}$ eV,
which is slightly larger than $E_{\rm p,\tau}$.
However, the synchrotron cooling time of muons is still shorter 
than the decay time scale for the relevant energy range. 
Thus, the conversion efficiency into high-energy photons is still large 
in the Continuous Model, too. 

\subsection{Electron--Positron Pair Production}

As was argued in subsection 2.3, a large fraction of  
the energy of accelerated protons is converted into high-energy photons 
both in the Discrete and Continuous Models. 
Typical energies of photons from neutral pions are above $10^{13}$eV 
while those from charged pions and muons are below $10^9$eV. 
The high-energy photons emitted from $\pi^0$s and $\pi^\pm$s
create electron--positron pairs through collisions with residual 
thermal photons.
The cross section of the pair creation is written as
$\sigma_\pm=\sigma_{\rm T} g(y)$
\citep{ber82},
where
%%%%%%%%%%%%%%%%%%%%%%%
\begin{eqnarray}
g(y)=\frac{3}{16} (1-y^2) \left[
(3-y^4) \ln{\frac{1+y}{1-y}}-2 y (2-y^2) \right].
\end{eqnarray}
%%%%%%%%%%%%%%%%%%%%%%%
The dimensionless value $y$ is defined by 
$y^2=1-(2 m_{\rm e}^2 c^4)/[\varepsilon_\gamma \varepsilon_\gamma' 
(1-\cos{\theta})]$.
In figure 6, we plot the time scale and the optical depth 
($\tau_{\gamma \gamma}$)
against the pair creation in the 100 eV radiation field 
for the Discrete Model.
Figure 6 shows that the photons with energy from 
$\varepsilon_{\gamma} = 3 \times 10^8$ eV
to $4 \times 10^{14}$ eV  
create electron--positron pairs before they escape from the shell 
with $l=10^9$ cm.
We define the threshold energy $\varepsilon_\tau$, above which
photons create pairs.
Photons from the decay of 
neutral pions and high-energy part of the synchrotron radiation from 
charged pions can be converted into pairs, while low-energy 
synchrotron photons will directly escape.  

\begin{figure}
\begin{center}
\FigureFile(100mm,170mm){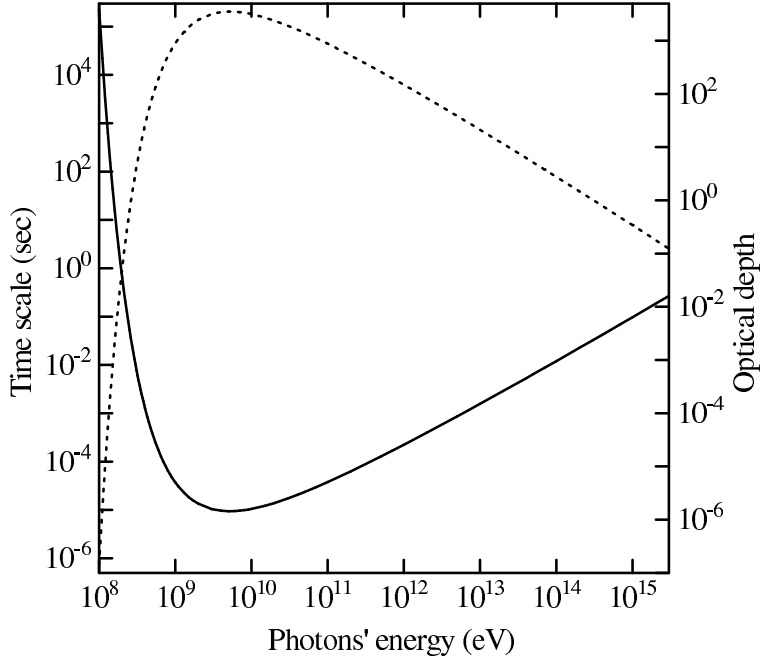}
\end{center}
\caption{Time scale (solid line) and the optical depth
(dotted line) of photons for the pair creation in the thermal photon field.
We assume $T=100$ eV and $l=10^9$ cm in this figure.
}
\end{figure}

In the Continuous Model, $\varepsilon_\tau$ turns out to be similar, 
since the larger available size compensates for the smaller photon 
density, which results in a similar value of $\tau_{\gamma \gamma}$.
Since both models give a similar input of high-energy photons, 
and a similar optical thickness to the pair absorption, 
the subsequent development of the cascade is also similar.   

First we consider pair production of photons from decay of the 
neutral pions.
Such electrons and positrons emit the synchrotron radiation 
the energy of which is given by 
%%%%%%%%%%%%%%%%%%%%%%%
\begin{equation}
\frac{\hbar q B}{m_{\rm e} c} 
\gamma_\pm^2
\sim 4 \left( \frac{B}{10^6 \mbox{G}} \right)
\left( \frac{\gamma_\pm m_{\rm e} c^2}{10^{10}{\rm eV}} \right)^2
 \quad \mbox{MeV},
\end{equation}
%%%%%%%%%%%%%%%%%%%%%%%
where $\gamma_\pm$ is the Lorentz factor of the pairs.
Since the quantum effect is important for electrons higher 
than $10^{13}$ eV for $B=10^6$ G,
the emitted energy is comparable to the energy of the emitter themselves.
Those photons create pairs again.
Reiterating the pair creation and the synchrotron radiation,
these high-energy photons cascade into low-energy pairs.

The energy of electrons/positrons which emit photons with 
a critical energy of $\varepsilon_\tau = 3 \times 10^8$ eV 
is $8 \times 10^{10}$ eV. 
Thus, the synchrotron photons emitted from electron/positrons 
above $8 \times 10^{10}$ eV are absorbed through pair production, 
while those emitted by pairs below $8 \times 10^{10}$ eV escape 
from the shell.
Thus, for $m_{\rm e} c^2 \gamma_\pm \geq
\varepsilon_\tau=3\times 10^8$ eV, the injection of 
pairs due to pair absorption of synchrotron photons emitted from 
higher energy pairs balancing with synchrotron cooling makes 
the resultant number spectrum of pairs as steep as 
$N_{\pm} \propto \gamma_\pm^{-3}$. 
On the other hand, below $3\times 10^8$eV, pair injection becomes 
unimportant, and the number spectrum of pairs is determined only by 
radiative cooling as $N_{\pm} \propto \gamma_\pm^{-2}$. 
Then, the typical energy of the synchrotron emission from pairs 
with break energy $\varepsilon_\tau=3 \times 10^8$eV is
%%%%%%%%%%%%%%%%%%%%%%%
\begin{equation}
\varepsilon_{\pm}=\frac{\hbar q B}{m_{\rm e} c} \left(
\frac{\varepsilon_{\tau}}{m_{\rm e} c^2} \right)^2
\sim 4 \left( \frac{B}{10^6 \mbox{G}} \right) \quad \mbox{keV}.
\label{ebr}
\end{equation}
%%%%%%%%%%%%%%%%%%%%%%%
As a result, the photon index of synchrotron radiation
becomes $-2$ above $\varepsilon_{\pm}$, while below that the 
photon index becomes $-1.5$. 
Since the energy of photons will be blue-shifted by
the relativistic motion of the shocked shell 
[$\Gamma_{\rm m} \simeq 
\Gamma_{\rm r} (\Gamma_{\rm R}-\sqrt{\Gamma_{\rm R}^2-1}) \sim 200$], 
this typical energy is observed as a few hundred keV, 
which is consistent with the observations.

In the Continuous Model, similar processes are generated, and 
the main differences lie in the behavior of $B$ and $\tau_{\gamma\gamma}$ 
as a function of $r$. 
In figure 7 we plot the typical photon energy $\varepsilon_{\pm}$ 
emitted from the pairs of typical energy $\varepsilon_{\tau}$ against $r$.
Here, we have assumed $\eta_2=1$, $\Delta_7=1$, and $f_{\rm B}=0.1$.
The value $\varepsilon_{\pm}$ has been calculated for radii
where the photopion production can occur by the thermal radiation field.
Since $\varepsilon_{\tau}$ does not change very much against the change in
the optical depth, $\varepsilon_{\pm}$ practically behaves as 
$\propto B \propto r^{-1}$ in figure 7.
The value $\varepsilon_{\pm}$ is around 0.1--10 keV in the wind frame,
which is typically observed as $0.01$--$1$ MeV, similar to 
the Discrete Model.
Thus, the results in the Discrete and Continuous Models are basically the same. 
In both models the gamma-ray emission can be ignited by
photopion production against the residual radiation field.

\begin{figure}
\begin{center}
\FigureFile(50mm,70mm){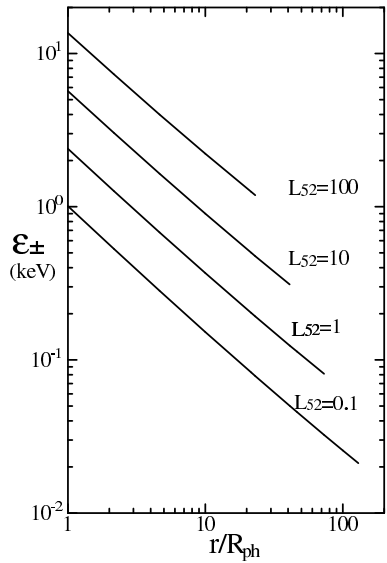}
\end{center}
\caption{Photon energy $\varepsilon_\pm$ in the comoving frame
for $\eta_2=1$ and $\Delta_7=1$ in Continuous Case.
}
\end{figure}

Next, we consider the fate of synchrotron photons emitted by 
charged pions and muons. Since the energy range is typically 
$10^5$--$10^9$ eV, only the high-energy portion of the spectrum 
is absorbed and produce pairs at around 
$\varepsilon_\tau=3 \times 10^8$ eV,
while the lower energy photons directly escape. 
Thus, around the break energy, synchrotron radiation of pairs 
originating from charged pions superposes on the smoother component 
of cascades from neutral pions. 

It is to be noted that this is not the whole story because the thus-produced photons
overweigh the residual photons and modify the cascading 
behavior itself. Thus, we expect that the optical thickness to the pair 
absorption drastically increases and that lower energy photons which have 
been regarded to directly escape in the above will also be pair-absorbed 
to enhance the cascade process.

\section{Emission from Cascading Particles}

\indent

As discussed in section 2, the residual radiation field 
of the fireball can trigger a cascade from shock-accelerated 
relativistic protons. The photon field from the cascading particles 
itself will overwhelm the thermal radiation field as the cascade process 
goes on, and the cascade process proceeds nonlinearly.
The energy density of the photons generated in the cascades is 
larger than that of the residual photons by about one order of magnitude. 
In this section we discuss the effects of the
self-generated photon field on the resultant emission spectra.

\subsection{Self-Generated Photon Field}

Self-generated photons drastically decrease the threshold energy 
of pair creation ($\varepsilon_{\tau}$) and make inverse Compton scattering 
as important as the synchrotron radiation as the cooling processes of pairs. 
The cascade process proceeds as shown in figure 8.
Although the whole process is time dependent, we here treat effects of 
these photons within the steady state one-zone model. 
In the Discrete Model, for simplicity, we assume that the fraction $f_\gamma$ of
the internal energy $U_{\rm in}$
in the shocked shell goes into the radiation field,
and the photon number density obeys a power-law distribution as
%%%%%%%%%%%%%%%%%%%%%%%
\begin{eqnarray}
n(\varepsilon_\gamma)=
(\alpha-2) \frac{f_{\gamma} U_{\rm in}}{\varepsilon_0^2}
\left\{
\begin{array}{l}
\left(\varepsilon_\gamma/\varepsilon_0 \right)^{-\alpha} \quad
({\rm for} \quad \varepsilon_\gamma >  \varepsilon_0) \\
0 \quad
({\rm for} \quad \varepsilon_\gamma \le  \varepsilon_0).
\\
\end{array}
\right.
\label{phfield}
\end{eqnarray}
%%%%%%%%%%%%%%%%%%%%%%%
Of course, this photon spectrum will not generally be coincident with
the radiation field that is finally realized.
As shown below, however, the typical energy of the cascading 
particles is not very sensitive to the shape of the photon spectra, but is
determined mainly by $f_\gamma$.

In the Continuous Model we assume that the photon density is similarly obtained
using $U$ in equation (\ref{U}).

\begin{figure}
\begin{center}
\FigureFile(100mm,170mm){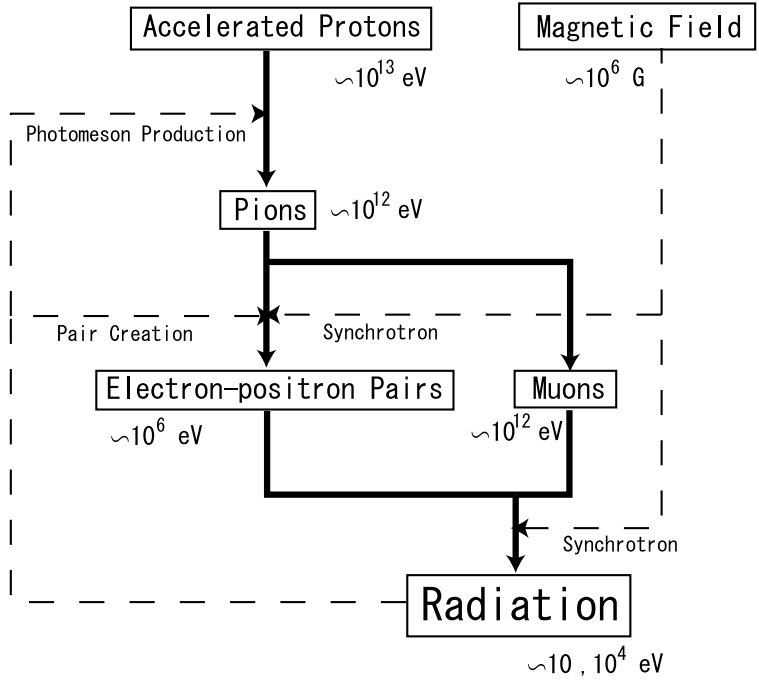}
\end{center}
\caption{Emission process when
the photon field due to the cascading particles itself overwhelms
the residual radiation (see section 3).
The notation is the same as in figure 1.
}
\end{figure}

\subsection{Photopion Production}

For the photon field given by equation (\ref{phfield}), $\tau_{\pi}$ is 
analytically obtained as
%%%%%%%%%%%%%%%%%%%%%%%
\begin{eqnarray}
\tau_{\pi}(\gamma_{\rm p})
= C f_{\gamma} 2^\alpha \frac{\alpha-2}{\alpha+1}
\frac{\sigma_{\rm eff}(\alpha)}{\sigma_{\rm T}}
\frac{m_{\rm p} c^2}{\varepsilon_0}
\left( \frac{\varepsilon_0 \gamma_{\rm p}}{m_{\rm e} c^2} \right)^{\alpha-1}
%\mathcal{R}^{-2}
,
\end{eqnarray}
%%%%%%%%%%%%%%%%%%%%%%%
where $\sigma_{\rm eff}(\alpha) \equiv 
\int d \chi \chi^{-\alpha} \sigma_{\pi}(\chi)$ and 
$\sigma_{\rm eff}(2.2) \simeq 2 \times 10^{-31} {\rm cm^2}$.
The dimensionless factor $C$ for the Discrete and Continuous Models
becomes $\Gamma_{\rm R}-1$ and $\mathcal{R}^{-1}$, respectively.
It is noted that the expression is independent of the parameters 
$M$ and $\Delta$ etc., because we have normalized the physical values
at $r=R_{\rm ph}$.

Since the threshold energy $E_{\rm p,\tau}$ for which $\tau_\pi=1$
has a dependence of 
$E_{\rm p,\tau} \propto \varepsilon_0^{(2-\alpha)/(\alpha-1)}$,
it turns out to be almost independent of $\varepsilon_0$ 
as long as $\alpha \simeq 2$.
For $\alpha=2.2$ and $\varepsilon_0=1$ keV,
all protons for which energy is larger than
%%%%%%%%%%%%%%%%%%%%%%%
\begin{eqnarray}
E_{\rm p,\tau}=
10^{13} \left( \frac{C f_{\gamma}}{0.3} \right)^{-5/6}
%\mathcal{R}^{5/3}
\quad {\rm eV},
\label{br1}
\end{eqnarray}
%%%%%%%%%%%%%%%%%%%%%%%
produce photopions. The threshold energy becomes an order of magnitude 
lower than in the previous section. Thus, the self-generated power-law 
photon field enhances the 
energy transfer from accelerated protons to photons and leptons. 
Since the inelasticity is about 0.2, the energy range of produced pions 
becomes $10^{12}$--$10^{16}$ eV. 

\subsection{Photon-Photon Pair Absorption}

The optical depth to the pair creation for a photon with energy 
$\varepsilon_\gamma$ in the photon field given by equation (\ref{phfield}) 
is analytically obtained as
%%%%%%%%%%%%%%%%%%%%%%%
\begin{eqnarray}
\tau_{\gamma \gamma}(\varepsilon_\gamma)
= C f_{\gamma} F(\alpha) \frac{m_{\rm p} c^2}{\varepsilon_0}
\left( \frac{\varepsilon_0 \varepsilon_\gamma}{m_{\rm e}^2 c^4} 
\right)^{\alpha-1},
%\mathcal{R}^{-2},
\end{eqnarray}
%%%%%%%%%%%%%%%%%%%%%%%
where
%%%%%%%%%%%%%%%%%%%%%%%
\begin{eqnarray}
F(\alpha) =\frac{4 (\alpha-2)}{\alpha+1}
\int_0^1 dy y (1-y^2)^{\alpha-2} g(y)
\end{eqnarray}
%%%%%%%%%%%%%%%%%%%%%%%
is a dimensionless function [$F(2.2) \simeq 2 \times 10^{-2}$].

If $f_\gamma \simeq 0.3$,
photons with energy $\gtrsim 1$ MeV suffer from pair absorption,
and do not directly escape from the shell.
Thus, the pair cascade 
fully develops over the whole energy region of relativistic pairs.
Synchrotron photons emitted from relativistic pairs distribute 
continuously over a wide energy range down to the $\sim 0.01$ eV
without any significant breaks. In the observer frame this energy is 
blue-shifted to a few eV, which is much less than the 
observed break energy.
In the Continuous Model, since $\tau_{\gamma \gamma} \propto r^{-1}$,  
the threshold energy $\varepsilon_\tau$ increases with $r$. 
However, the break energy of the synchrotron radiation does not become 
large enough because the magnetic field strength also 
decreases with $r$.
Even for $r=100 R_{\rm ph}$ with the fiducial parameters,
$\varepsilon_\tau \sim 10$ MeV and
$\varepsilon_\pm \sim 0.01$ eV in the comoving frame.

Before we discuss the detailed behavior of 
the cascade, we examine if photons from pions having energies 
above $\sim 1$ MeV. 

\subsection{Photons from Pions and Muons}

The produced $\pi^0$s promptly decay into two gamma-rays.
As is the case of the residual thermal photon field,
these photons with energy above $\sim 10^{12}$eV cascade 
into lower energy pairs.

As for the charged pions, high-energy pions cool through 
Compton and synchrotron radiation before decaying into muons, 
although Compton cooling is fairly suppressed by the Klein--Nishina 
effect in this energy range. 
In a similar way to that discussed in subsection 2.3,
we obtain the Lorentz factor of pions,
$\gamma_{\pi 0}$, whose lifetime is comparable to the cooling time as
%%%%%%%%%%%%%%%%%%%%%%%
\begin{eqnarray}
\gamma_{\pi 0}^2
=\frac{3 m_\pi c}{4 t_{\pi,0}}
\frac{\mathcal{L}}{(f_\gamma+f_{\rm B}) m_{\rm p} c^2}
\left( \frac{m_\pi}{m_{\rm e}} \right)^2,
\label{picool}
\end{eqnarray}
%%%%%%%%%%%%%%%%%%%%%%%
where the length scale $\mathcal{L}$ for the Discrete and Continuous Models
becomes $l/(\Gamma_{\rm R}-1)$ and $R_{\rm ph} \mathcal{R}^2/\eta$, respectively.
We should note that
the typical value of $\mathcal{L}$ ($\sim 10^{11}$ cm)
in the Continuous Model 
is much larger than that ($\sim 10^9$ cm) in the Discrete Model.
The pions, whose energy is larger than
$\gamma_{\pi 0} m_\pi c^2 \simeq 2 \times 10^{13} (\mathcal{L}/10^9 {\rm cm})^{1/2}
[(f_\gamma+f_{\rm B})/0.4]^{-1/2}$ eV,
should cool down before they decay into muons and neutrinos.
According to equation (\ref{pi-syn}), synchrotron photons from charged pions 
above $3\times 10^{13}$eV suffer from pair absorption, while 
those from charged pions below $3\times 10^{13}$eV do not. 
Thus, a major portion of the energy of high-energy charged pions will 
be injected into the cascade process, and some part results in the 
emission of photons in the sub-MeV range. 

Although charged pions below $\sim 10^{13}$eV decay into muons and neutrinos before 
significant radiative cooling,
the muons take a large fraction ($\sim m_{\mu}/m_{\pi} \sim 0.8$) 
of the original energy of pions and contribute to photon emission.
Muons cool down through the synchrotron and inverse Compton mechanisms.
As is seen when we estimate $\gamma_{\pi 0}$,
it turns out that the muons, whose energy is larger than
$\gamma_{\mu 0} m_\mu c^2 \sim 10^{12} (\mathcal{L}/10^9 {\rm cm})^{1/2}
[(f_\gamma+f_{\rm B})/0.4]^{-1/2}$ eV,
should cool down before they decay into electrons.
Since the Klein--Nishina suppression works for target 
photons of energy larger than 10 keV,
we should take this limit into account for $f_\gamma$.
In the Discrete Model,
since muons are injected in the energy range of $\sim 10^{12}$--$10^{13}$ eV,
muons lose more than a half of the initial energy radiatively before decay. 
The cooling time is $t_{\mu, {\rm cool}} \simeq 0.03 (\mathcal{L}/10^9 {\rm cm})^{1/2}
[(f_\gamma+f_{\rm B})/0.4]^{-1/2}$
s that is comparable to the dynamical time scale $l/c$.

The typical energy of muon synchrotron radiation is
%%%%%%%%%%%%%%%%%%%%%%%
\begin{eqnarray}
\frac{\hbar q B}{m_\mu c} \gamma_{\mu 0}^2
&=& \frac{3 \sqrt{2 \pi} \hbar q}{2 t_{\mu,0}} \frac{f_{\rm B}^{1/2}}
{f_\gamma+f_{\rm B}}
\sqrt{\frac{\mathcal{L}}{m_{\rm p} c^2 \sigma_{\rm T}}} \\
& \simeq & 20 \left( \frac{f_{\rm B}}{0.1} \right)^{1/2}
\left( \frac{f_\gamma+f_{\rm B}}{0.4} \right)^{-1}
\left( \frac{\mathcal{L}}{10^9 {\rm cm}} \right)^{1/2} 
%\mathcal{R}
 \quad {\rm keV}.
\end{eqnarray}
%%%%%%%%%%%%%%%%%%%%%%%
This energy should be blue-shifted by 
the relativistic motion of the shocked shell ($\Gamma_{\rm m} \sim 200$)
and observed as $\sim$ MeV, which is slightly higher than the observed break 
energy of GRBs.  
The synchrotron spectrum from muon cooling is 
broadly concentrated around the above typical energy. 
The muons, whose energy is smaller than
$\gamma_{\mu 0} m_\mu c^2$, decay into electrons.
These electrons emit very high-energy photons
that cannot escape from the shell because of pair absorption.

In the Continuous Model, since $\mathcal{L} \sim 10^{11}$ cm,
the typical energy of muon synchrotron becomes an order of magnitude
larger than in the Discrete Model.

\subsection{Pair Cascade}

In the cascade process, as long as the energy density of
power law photons is comparable to that of the
magnetic field, the Compton scattering also produces high-energy photons,
although Klein--Nishina suppression is significant in the high-energy regime. 
As the energy of pairs in the cascade process decreases, 
the Compton scattering plays a major role.
In addition, created pairs increase the optical thickness to the scattering.
These processes may modify the resultant spectrum.
In order to obtain the detailed spectrum in the pair-cascade model
we need to simulate electron--positron plasmas
including the effects of synchrotron radiation, Compton scattering, 
pair creation, pair annihilation, Coulomb scattering, e--e bremsstrahlung, 
and so on.
However, such numerical calculation is beyond the scope of this paper.

Here, we qualitatively discuss the spectrum emitted from cascading pairs,
especially the possible breaks in the spectrum.
We simply estimate the prospective physical conditions of the plasmas
with the help of published numerical simulations made for pair cascade models
for compact X-ray sources \citep{gui83,lig87,cop92,ste95}.
These previous simulations assumed that 
copious soft photons are injected at lower energies,  
which correspond to the residual thermal photons of the fireball 
in the present case. In the present case, the magnetic field is relatively 
large; its energy density is comparable to that of injected high-energy 
photons.   

In terms of the compactness parameter, our case corresponds to 
%%%%%%%%%%%%%%%%%%
\begin{equation}
l_{\rm e}\equiv\frac{L_{\pi}\sigma_{\rm T}}{l m_{\rm e}c^3} 
=C f_\pm
\frac{m_{\rm p}}{m_{\rm e}} \simeq 550 \left( \frac{C f_\pm}{0.3} \right), 
%\mathcal{R}^{-2},
\end{equation}
%%%%%%%%%%%%%%%%%%%%%%%%%%
where we have taken a volume $l^3$ because our case is not spherical.
Although this typical value of the compactness parameter is somewhat 
large compared with those adopted in the above-mentioned studies, the qualitative 
features will not be different. 

The created pairs above $\sim 1$ MeV will distribute in a 
monotonic fashion.
The pairs cool down mainly through synchrotron radiation and
inverse Compton scattering to form a power-law spectrum with a number index of
around 3, as discussed in subsection 2.4.
This corresponds to the case when all of the 
injected energy flows down in the energy space while making lower 
energy pairs and photons. 
The number spectra of synchrotron emission and inverse Compton 
scattering is described by a power law of index around 2. 
This very rough consideration is a fairly good approximation 
as the simulations with the power law injection of pairs
by \citet{cop92} showed the power law indices of photons
in the case of $f_\gamma \sim f_{\rm B}$ become around 2 
in the range $x \equiv \varepsilon_\gamma/m_{\rm e} c^2
=10^{-3}$--$10^{-1}$, which we are interested in.
Below this energy, the effects of cooled pairs appear while above $x=1$ 
the effects of pair absorption play a role.

As the low-energy pairs accumulate, a sharp break due to the synchrotron 
self-absorption appears
in the spectrum in the case of $f_{\rm B} \simeq f_\gamma$.
Assuming $N_{\pm} \propto \gamma_\pm^{-(p+1)}$ above $E_{\rm \pm, min}$,
we obtain the synchrotron self-absorption frequency \citep{ryb79} as
%%%%%%%%%%%%%%%%%%%%%%%
\begin{eqnarray}
\varepsilon_{\rm abs} \simeq 10
\left( \frac{f_\gamma+f_{\rm B}}{0.4} \right)^{-0.28}
\left( \frac{f_{\pm}}{0.3} \right)^{0.28} 
\left( \frac{f_{\rm B}}{0.1} \right)^{0.36}
\left( \frac{\mathcal{L}}{10^9 {\rm cm}} \right)^{-0.36}
\left( \frac{E_{\rm \pm, min}}{1 {\rm MeV}} \right)^{0.61}
%\mathcal{R}^{-0.72}
\quad {\rm eV},
\end{eqnarray}
%%%%%%%%%%%%%%%%%%%%%%%
for $p=2.2$.
The predicted self-absorption energy becomes $\sim 2$ keV in 
the observer frame for $\Gamma_{\rm m} \simeq 200$, lower than 
the observed break energy.

Next we discuss the cooled pairs. 
Following the convention, we define 
the ``pair yield'' as 
$PY \equiv \int d \gamma_\pm P(\gamma_\pm)
          /\int d \gamma_\pm Q(\gamma_\pm)(\gamma_\pm-1)$, 
where $Q(\gamma_\pm)$ is the rate of injection of pairs of 
$\gamma_\pm$ via the pion decay,
and $P(\gamma_\pm)$ is the rate at which secondary pairs of $\gamma_\pm$
are created via pair production.
Since the numerical simulations indicated that for 
$l_{\rm e} \gg 10$ the value $PY$ becomes $\sim 0.1$ independent
of other parameters, 
we assume $PY \simeq 0.1$ hereafter. This means that 10\% of the 
injected energy goes into the rest mass energy of cooled pairs 
and the remaining 90\% of energy goes into radiation. 
Created pairs may cool down to subrelativistic energies and thermalize
before annihilating.
In the steady state, the pair creation rate is assumed to balance 
with the rate of pair annihilation.
We thus obtain
%%%%%%%%%%%%%%%%%%
\begin{equation}
\frac{3}{8} \sigma_{\rm T} c n_+ n_- =
\frac{1}{2}\int P(\gamma_\pm) d\gamma_\pm,
\end{equation}
%%%%%%%%%%%%%%%%%%%%%%%%%%
where $n_+$ and $n_-$ are the number densities of cooled thermal positrons
and electrons, respectively.
The Thomson optical depth to thermal pairs is written by
%%%%%%%%%%%%%%%%%%
\begin{equation}
\tau_\pm=\left( \frac{4}{\pi} l_{\rm e} PY \right)^{1/2}=7 
\left( \frac{C f_\pm}{0.3}
\right)^{1/2} \left( \frac{PY}{0.1} \right)^{1/2}.
%\mathcal{R}^{-1}.
\label{taupm}
\end{equation}
%%%%%%%%%%%%%%%%%%%%%%%%%%
The equilibrium temperature of the thermal pairs, $T_\pm$,
is mainly determined by requiring that no net energy be transferred between
the particles and photons.
Numerical simulations for compact X-ray sources in the past 
showed $\Theta \equiv T_\pm/m_{\rm e} c^2 \sim 10^{-2}$--$10^{-3}$ for $l_{\rm e} \gg 10$
regardless of the other parameters.
Therefore, the Compton $y$ parameter, $y \sim 4 \Theta \tau_\pm^2 \lesssim 1$, 
in our model and relatively mild Comptonization may occur. 
The photon energy tends to be peaked around at 1--10 keV, 
which is expected for the observed break energy of GRBs when 
beaming effects are taken into account. 

Because of the large Thomson optical depth of thermal pairs 
($\tau_\pm$), the mean energy loss of photons with energy lower than 
$m_{\rm e}c^2$ in one scattering by cold thermal pairs is about 
$\varepsilon_\gamma^2/m_{\rm e} c^2$,
and the mean number of scattering is about $\tau_{\pm}^2$.
A significant depletion of the spectrum may occur in the region
$1/\tau_{\pm}^2 \lesssim x \lesssim 1$ \citep{lig87,cop92,ste95}, 
due to downscattering
of photons by cold thermal pairs.
The break energy, $m_{\rm e} c^2/\tau_{\pm}^2  \simeq$ a few or a few 
tens of keV, can be the same
order of observed break energy owing to the relativistic blue-shift.
Thus, cooled thermal pairs may possibly produce spectral break 
compatible with observations through thermal Comptonization and 
downscattering of high-energy photons. 
However, in order to quantitatively investigate this possibility,
detailed numerical simulations will be required. 

\section{Discussion} 

We have shown that Fermi accelerated protons with $\gtrsim 10^{13}$ eV 
efficiently bring their energy into pion production and subsequent 
photon emission.
The photon spectra of the cascades from 
relativistic protons are mainly characterized by synchrotron 
and inverse Compton components emitted from pairs. In addition, muon 
synchrotron radiation can make an important contribution.
Cooled thermal pairs can modify the photon spectra through 
Comptonization and downscattering of high-energy photons. 
In figure 9, we show the schematic spectra which we discussed above.
Although we have discussed several mechanisms to reproduce the observed 
break energy scale including muon synchrotron radiation, 
synchrotron self-absorption, Comptonization, and down scattering 
by cooled thermal pairs, at present we have not definitely identified 
any specific one with the observed break feature.
The problem remains to be open.

\begin{figure}
\begin{center}
\FigureFile(120mm,80mm){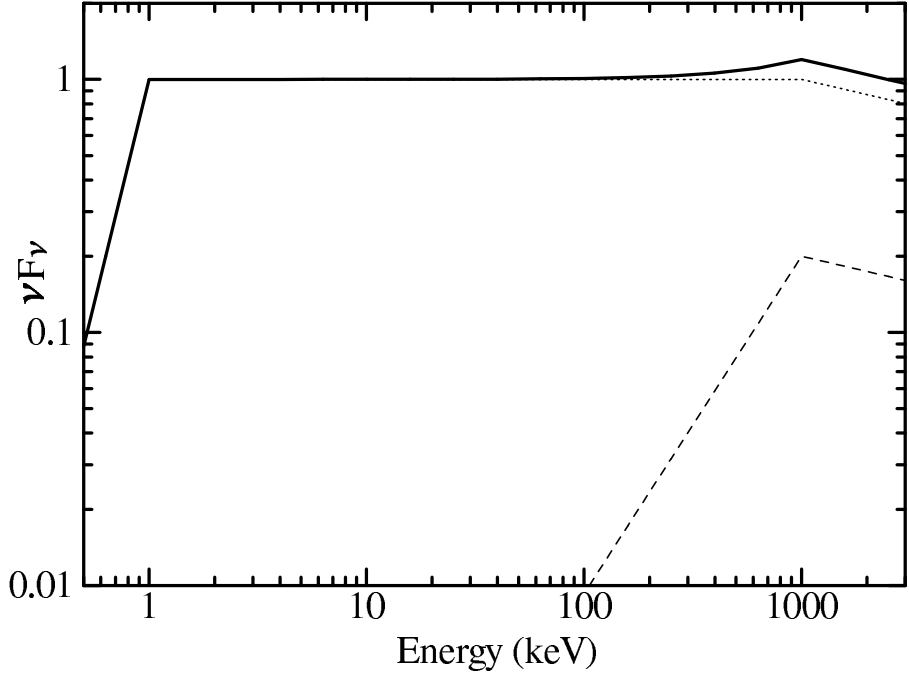}
\end{center}
\caption{Schematic photon spectrum in the observer frame.
The dotted and dashed lines are
the components due to the electron--positron pairs and muons, respectively.
The solid line shows the integrated spectrum.
Here, we assume that the energy of the component from the muons
is much smaller than the energy of the component from the pairs.
The break due to the self-absorption appears at $\sim 1$ keV.
The downscattering by cold thermal pairs makes
both the components above $\sim 1$ MeV decrease.
}
\end{figure}

Let us briefly discuss this problem.
The simplest way to produce a sharp spectral break
is to prevent electrons/positrons from cascading below the typical energy,
$\varepsilon_\tau$.
As discussed in section 2, if the photon density in the shocked
region is sufficiently low, the pair-creation process cannot
provide electron--positron pairs under energy $\varepsilon_\tau$.
The value of $\varepsilon_\tau$ should be
large enough ($\sim 100 m_{\rm e} c^2$)
to reproduce the spectral break [see equation (\ref{ebr})].
However, the condition that the emission from the cascading particles
themselves is more luminous than the residual radiation
of the fireball implies $f_{\gamma} \gtrsim 0.1$ consistent
with the value we have adopted in section 3.
In addition,
$f_{\gamma}$ may be larger than $\sim 0.1$ to reproduce the energetics of GRBs.
Thus, the condition $f_{\gamma} \gtrsim 0.1$ is universally required
so that the pair cascade develops over the whole energy region of relativistic pairs
(namely $\varepsilon_\tau \sim m_{\rm e} c^2$).

As shown in figures 5 and 7, in the Continuous Model
the cascade process can occur even far outside of $R_{\rm ph}$.
If internal shocks occur far beyond $R_{\rm ph}$,
the photon density is diluted and
$\varepsilon_\tau$ can be large.
As discussed in subsection 3.3, however,
the smaller magnetic field prevents the break energy
from increasing sufficiently.

We have neglected effects of the directly accelerated electrons.
If the total energy of electrons directly accelerated
is more than a tenth of the energy of the protons,
the synchrotron emission from the electrons can make
the ignition of photopion production easier.
However, the emission from the electrons does not modify
the cascade process very much.
Therefore, even in this case
$\varepsilon_\tau$ remains too small to reproduce the observed 
break energy.

One may think that the cascade process can be largely altered if $f_{\rm B}
\ll 0.1$.
Here, we argue that this is not the case.
As far as $B>100$ G with $f_{\rm B} > \sim 10^{-9}$, protons can be accelerated
beyond $10^{13}$ eV for $l=10^9$ cm and photopion production against
residual photons occurs efficiently.
While synchrotron cooling of charged pions is strongly suppressed, neutral
pions decay
into two gamma-rays and create electron--positron pairs through collisions
with residual thermal photons.
The energy of those pairs is comparable to the energy of $\pi^0$s
($\sim 10^{13}$ eV). Even though the Klein--Nishina suppression makes
inverse Compton process ineffective for such pairs,
the pairs cool down through synchrotron radiation;
even for $B=100$ G the cooling time of pairs via synchrotron radiation is
shorter than the dynamical time scale because of their large Lorentz factor.
The energy of synchrotron photons emitted from the pairs
is larger than 100 MeV for $B \ge 100$ G.
Therefore, the synchrotron photons create secondary pairs again (see subsection
2.4)
and the energy of $\pi^0$s is transfered to lower energy pairs
via the cascade process illustrated above.
The low energy pairs cool through inverse Compton process against
residual photons and produce the self-generated photon field.
When the self-generated photons overwhelm the residual photons
($f_{\gamma} \gtrsim 0.01$), the cooling processes of low-energy pairs,
charged pions, and muons are dominated by inverse Compton process,
rather than synchrotron radiation.
For $f_{\gamma} \gtrsim 0.1$ the values of $\gamma_{\pi 0}$
[see equation (\ref{picool})], $\gamma_{\mu 0}$, and $\varepsilon_\tau$
remain close to the values for $f_{\rm B}=0.1$.
Thus, even when $f_{\rm B} \ll 0.1$,
the cascade process proceeds in a similar way as discussed in section 3.
Cooled muons with $\gamma_{\mu 0}$ decay and provide electrons/positrons
within a narrow energy range around $10^{11}$ eV (see subsection 3.4).
Photons emitted from such electrons/positrons can be in a narrow
energy spectrum, and escape without producing pairs.
Even for $B=100$ G, the energy of the synchrotron photons
emitted from the electrons of $10^{11}$ eV is 40 keV
in the comoving frame.
This energy is an order of magnitude larger than the typical break energy.
Since the Klein--Nishina suppression is significant for such electrons/positrons,
the cooling time also tends to be longer than the dynamical time scale.
In any case, most of the self-generated photons come from neutral pions.
The spectrum component due to the cascade process initiated from $\pi^0$-decay
may not reproduce the break energy.

From the above discussion,
pairs generally distribute over a wide energy range,
which overpredicts the flux of low-energy photons
between $\sim 1$ keV and $\sim 1$ MeV within a simple analysis.
There may be no physical process to prevent electrons/positrons from cascading
into lower energy.
The possible mechanisms to make the spectral break may be
Comptonization or downscattering of high-energy photons
by cooled thermal pairs.
It is difficult to estimate the precise effects of
Comptonization and downscattering analytically.
The Comptonization may resolve the problem of the spectral index below the break.
We have assumed $f_{\rm B} \sim f_\gamma$ in our discussion.
A smaller value of $f_{\rm B}$ makes inverse Compton scattering
dominant rather than synchrotron radiation.
Some numerical simulations showed that the spectral index decreases with
decreasing $f_{\rm B}/f_\gamma$ \citep{zdz86,cop92}.
A smaller value of $f_{\rm B}$ is favorable to decrease the spectral
index below the break.
After all, we need to simulate the cascade and emission processes numerically
to predict the final spectra quantitatively.

Our model may explain some puzzling features of GRBs.
For example, some spectra obtained by the Ginga satellite
show break energies of a few or a few ten keV and spectral number 
index smaller than 2
below the breaks \citep{str98}.
These break energies are close to the self-absorption frequency
in our model.
Since the emission spectra in the X-ray and gamma-ray range 
should depend on details of the model parameters, such as the 
positions of the internal shock and photosphere, our 
model may provide a wide range of spectral and timing behavior. 
It is interesting to see if our model can reproduce not only the standard 
spectral feature, but also
the anomalous spectrum, like the ``X-ray rich GRB'' \citep{hei01,kip01}
or GRB precursor \citep{kos95}.

Our scenario also predicts several features of neutrino emission. 
The cooling of muons and pions results in a clustering of neutrino 
energies.
From the typical energy of pions at the decaying moment,
we obtain the energy of $\nu_{\mu}$ ($\bar{\nu}_{\mu}$), which is produced from 
the decay of
$\pi^+$ ($\pi^-$)
as $0.2 \gamma_{\pi 0} m_\pi c^2 \Gamma_{\rm m} \simeq 9 \times 10^{14} 
(\Gamma_{\rm m}/200)$ eV in the observer frame.
If $f_\gamma$ and $f_{\rm B}$ are sufficiently small, the energy of neutrinos 
can be large.
In this case, however, a small $f_\gamma$ means a weak flux of neutrinos.
The energy of $\nu_{\rm e}$ and $\bar{\nu}_{\mu}$ 
($\bar{\nu}_{\rm e}$
and $\nu_{\mu}$), which is produced from the decay of
$\mu^+$ ($\mu^-$),
is as small as $\sim 7 \times 10^{13} (\Gamma_{\rm m}/200)  
%\mathcal{R}
$
eV in the observer frame.

\section{Conclusions}

\indent

We have investigated the consequences of proton acceleration 
in the residual thermal photon field in the internal shocks of GRBs. 
Even if the internal energy of heated protons cannot
be converted into the energy of electrons,
radiation energy can be drained from the accelerated protons
through the photopion production process against residual 
thermal photons.
We have examined two extreme models concerning the distribution of 
residual thermal photons. In the Discrete Model, photons are 
confined within multiple thin shells, while in the Continuous Model, 
photons are emitted from the wind photosphere. 
In both the Discrete and Continuous Models
the residual radiation field in the fireball is shown to be able to ignite 
the photopion production and the generated photon field further 
enhances the photon-initiated cascade.
Although the real situation lies between these two extreme models,
there are no large differences in the ignition processes.
The Fermi-accelerated protons with $\gtrsim 10^{13}$ eV
bring their energy into pions.
If the number fraction $\zeta$ of protons are accelerated and
distributed as $n_{\rm p} \propto E_{\rm p}^{-q}$, 
and if they have the fraction $\xi$ of the internal energy,
the energy fraction of pions is obtained by
%%%%%%%%%%%%%%%%%%%%%%%
\begin{eqnarray}
f_\pi&\simeq&\left( \frac{q-2}{q-1} \right)^{q-2} \zeta^{2-q} \xi^{q-1}
\left( \frac{E_{\rm p,\tau}}{m_{\rm p} c^2} \right)^{2-q} \\
&=& 0.3 \left( \frac{\zeta}{0.01 \%} \right)^{-0.2} \left( \frac{\xi}{50 \%} \right)^{1.2}
\left( \frac{E_{\rm p,\tau}}{10^{13} {\rm eV}} \right)^{-0.2},
\end{eqnarray}
%%%%%%%%%%%%%%%%%%%%%%%
for $q=2.2$.

The particles cascading from the pions emit photons over a wide energy range.
The emission properties for the two models are also similar.
The spectrum is composed of mainly two components:
emissions from the electron--positron pairs and muons,
produced by high-energy photons and charged pions, respectively.
The pairs form a power-law spectrum with a number index of around 3
over the whole energy region of relativistic pairs.
On the other hand muons cool down to a characteristic energy
before they decay.
The radiation due to the pairs 
is composed of a power-law spectrum with an index of about 2
with possible break features around the 1--10 keV 
range in the comoving frame
due to Comptonization and downscattering by cooled thermal pairs. 
This may correspond to the observed break when the relativistic 
beaming effects are taken into account. 
The muons can emit X and gamma-rays by synchrotron process and produce
a sharp break at around 1--10 MeV in the spectrum.

\vspace{1.5cm}

This work is  supported in part by a Grant-in-Aid for Scientific Research 
from the Japanese Ministry of Education, Culture, Sports, Science and
Technology (No.13440061, F.T.).
One of the authors (K.A.) was supported by the Japan Society for the Promotion of Science.

\end{document}